# Reaction of O$_2$ with Subsurface Oxygen Vacancies on TiO$_2$ Anatase (101)

Martin Setvín,[1] Ulrich Aschauer,[2†] Philipp Scheiber,[1] Ye-Fei Li,[2] Weiyi Hou,[2] Michael Schmid,[1] Annabella Selloni,[2] and Ulrike Diebold[1*]

[1]Institute of Applied Physics, Vienna University of Technology, Wiedner Hauptstrasse 8-10/134, 1040 Vienna, Austria.

[2]Department of Chemistry, Princeton University, Frick Laboratory, Princeton NJ 08544, USA.

*Correspondence to: Ulrike Diebold (diebold@iap.tuwien.ac.at)

[†] Present address: ETH Zürich, Materials Theory, Wolfgang-Pauli-Strasse 27, 8093 Zürich, Switzerland

**Abstract**:

Oxygen adsorbed on metal oxides is important in catalytic oxidation reactions, chemical sensing, and photocatalysis. Strong adsorption requires transfer of negative charge from, e.g., oxygen vacancies or dopants. With scanning tunneling microscopy (STM) we observed, transformed, and, in conjunction with theory, identified the nature of O$_2$ molecules on the (101) surface of anatase (titanium oxide) doped with niobium. Oxygen vacancies (V$_O$'s) reside exclusively in the bulk, but we pull them to the surface with a strongly negatively charged STM tip. O$_2$ adsorbed as superoxo (O$_2^-$) at fivefold Ti sites was transformed to peroxo (O$_2^{2-}$), and, via reaction with a V$_O$, placed into an anion surface lattice site as an (O$_2$)$_O$ species. This so-called bridging dimer also formed when O$_2$ directly reacted with (sub-)surface V$_O$'s.

**One Sentence Summary:** O$_2$ adsorbed on anatase TiO$_2$ merges with subsurface oxygen vacancies to form a bridging O$_2$ dimer; a novel O$_2$ species that has been predicted theoretically to form at oxide surfaces.

**Main Text:**

Molecular O$_2$ interacts weakly with fully-oxidized metal oxides. When excess electrons are present, it adsorbs as an anion in either superoxo (O$_2^-$), peroxo (O$_2^{2-}$), or dissociated (2×O$^{2-}$) form. The negative charge can be provided by intrinsic defects (e.g., oxygen vacancies (V$_O$) or cation interstitials) in a reduced oxide, by doping, or through photo-excitation. Such adsorbed oxygen plays a key role in several technological processes, e.g., in oxidation reactions promoted by heterogeneous catalysts, in gas sensors, or in photocatalysis (*1, 2*). Because of the complexity of technical materials, multiple oxidation states, and the background of lattice oxygen, a molecular-level understanding of O$_2$ adsorption is just emerging (*3, 4*). STM studies on macroscopic, single-crystalline samples with flat surfaces and controlled defects allow a direct view of adsorbed species. In conjunction with density functional theory (DFT) calculations, they provide fundamental insights into a rich surface chemistry (*3-9*). However, experimental studies have not focused on anatase that makes up nanophase titania, nor have they addressed the role of subsurface V$_O$'s that are prevalent in this material (*10, 11*). Here we show that manipulations with the STM tip can be used to (i) locally create surface V$_O$'s, (ii) alter the charge state of adsorbed O$_2$, and (iii) react adsorbed O$_2$ with subsurface V$_O$'s to produce an interstitial (O$_2$)$_O$ species [or bridging O$_2$ dimer] that has been predicted theoretically.

Our sample was an anatase mineral single crystal, naturally doped with 1.1 at% niobium [see the Supplemental Online Materials (*12*)], an efficient electron donor. Although the thermodynamically stable rutile phase [especially the (110) surface of single crystals] has been extensively used as a model surface (*3, 4, 13*), nanophase TiO$_2$, which is often used in applications, usually consists of anatase (*14*). Anatase is also frequently referred to as 'photocatalytically more active', and several studies (*15-17*) have shown that substantial amounts of $O_2$ are present on the surface of photo-irradiated anatase, but not on rutile. This could be caused by the different response to photo-excitation, notably by long-lived photoexcited electrons present in anatase but not in rutile (*18*), which provide the charge for oxygen adsorption, or by the specific surface properties of anatase (*2, 19*).

In contrast to $TiO_2$ rutile (110), $V_O$'s are not stable at the surface of anatase (*10, 11*). When surface $V_O$'s are created by electron bombardment, they move into the bulk at temperatures as low as 200 K (*11*). In DFT calculations for an anatase slab with a subsurface $V_O$ (which provides negative charge), $O_2$ adsorbs molecularly at 5-coordinated surface $Ti_{5c}$ atoms (see Fig. 1A), as a peroxo $O_2^{2-}$ and a superoxo $O_2^-$ species at low and high coverages (*20*), respectively. Ti interstitials ($Ti_{int}$'s), which lead to dissociation of $O_2$ at rutile (110) (*5*), have the same effect on anatase (101) (*21*).

A peroxide $(O_2^{2-})_{ads}$ adsorbed in the vicinity of a subsurface $V_O$ (Fig. 1A) induces substantial structural relaxation, suggesting the existence of an energetically more favorable configuration. Indeed, in first-principles molecular dynamics (FPMD) simulations (at 220 K) we observed the cascade of events shown in Fig. 1, B to E. In the resulting structure, the $O_2$ takes the position of a twofold coordinated O ($O_{2c}$) atom in a bridging, 'side-on' $\eta^2$ configuration; in Kröger-Vink notation we refer to this species as an $(O_2)_O$. The $(O_2)_O$ retains a bond length of 1.46 Å, characteristic of $O_2^{2-}$ (*20*). This species (often referred to as 'interstitial' or 'bridging' $O_2$) has been predicted consistently in theoretical calculations of oxygen in/on anatase (*22, 23*), and proposed to be an important intermediate in the photocatalytic splitting of $H_2O$.

If such an $(O_2)_O$ indeed exists, then it should also form when an $O_2$ directly reacts with a surface $V_O$. To test this hypothesis, we prepared surface $V_O$'s on anatase (101) (Fig. 2). In previous work, we created such $V_O$'s by bombarding $TiO_2$ with electrons (*11, 24*). Here we find that identical defects can be generated by the STM tip. Fig. 2 shows STM images after scanning with high bias voltage and tunneling current. The bright spots within a well-defined area (Fig. 2A) are the same $V_O$'s as the ones generated by electron irradiation (*11*). After dosing with $O_2$ (Fig. 2) some of the $V_O$'s were replaced by double spots located at $O_{2c}$ sites. Their appearance agrees well with calculated STM images of the $(O_2)_O$ configuration (Fig. S4). The $(O_2)_O$ did not form at temperatures below 20 K, but it readily appeared when $O_2$ was dosed above 40 K, with a sticking coefficient $S$ near unity. This result roughly fits the DFT-derived activation energy of 47 meV.

When we dosed a surface without such artificially-created surface $V_O$'s, we also see the same $(O_2)_O$, albeit quite rarely (see Supplement). Most often, we observed two different $O_2$ species (Fig. 3). One of them, denoted $(O_2)_{extr}$, is related to the Nb dopant that preferably replaces a surface $Ti_{6c}$ atom (see Fig. S2 and Table S2). These dopants are distributed unevenly at the surface, with an average concentration of 0.5% of a monolayer (ML, defined as the density of surface $Ti_{5c}$ atoms), and a local variation between 0.1 and 1% ML. In STM, these impurities exhibit a typical triangular shape (see Figs. 3A, S1). Upon $O_2$ exposure at 105 K, bright dimers were found at the position of these extrinsic dopants; they adsorbed with $S \approx 0.1$. In DFT calculations (Fig. S3), the adsorption of an $O_2$ molecule at a $Ti_{5c}$ is preferred by 0.12-0.15 eV when next to a $Nb_{6c}$.

In addition to $(O_2)_{extr}$, a bright species, labeled $(O_2)_{ads}$, formed with with $S \approx 10^{-4} - 10^{-2}$ when anatase (101) was exposed to $O_2$ at 100 K. (The higher values were observed on a more reduced crystal.) The $(O_2)_{ads}$ is located at regular $Ti_{5c}$ surface atoms. Its concentration increases linearly with $O_2$ exposure (Fig. S5). The highest coverage observed was 5% ML. For higher exposures STM imaging becomes difficult; possibly the saturation coverage is significantly higher.

The $(O_2)_{ads}$ can be converted into other species with the STM tip. Fig. 3A shows several $(O_2)_{ads}$ and $(O_2)_{extr}$. During scanning at $V_{sample} = +3.2$ V (Fig. 3B) horizontal streaks indicate that the adsorbates were modified by the presence of the tip. (The slow scan direction was from bottom to top.) The $(O_2)_{ads}$ became more dimer-like (Fig. 3C) and the dark contours indicate that this new species carried a more negative charge. When subjecting these intermediate species to a slightly higher voltage (Fig. 3D), a second conversion took place, see Fig. 3E. The resulting species is the $(O_2)_O$ that also formed when an $O_2$ molecule directly reacted with a surface $V_O$. (See Supplement for more experimental evidence.) Also note that the $(O_2)_{extr}$ was converted into two $(O_2)_O$ already during the first high-voltage scan (Figs. 3, B and C), and that two $(O_2)_O$ were always produced per one $(O_2)_{extr}$, while only one $(O_2)_O$ resulted from each $(O_2)_{ads}$ (Fig. 3F). For example, the two neighboring $(O_2)_O$ pointed out in the initial image (Fig. 3A) resulted from previously scanning an $(O_2)_{extr}$ at high bias. The tip-induced conversion was reproducible on different samples and with different tips; with a minimum $V_{sample}$ was $+3.3 \pm 0.1$ V.

From these results, we derive a complete picture of the reaction of $O_2$ with a reduced $TiO_2$ anatase (101) surface. When $O_2$ is incident on the cold (~100 K) sample, it is weakly adsorbed. Stronger adsorption occurs when electrons are transferred from the surface to the molecule. The $Nb^{5+}$ dopants with nearby electrons are populated first, resulting in $(O_2)_{extr}$ species. A molecule can also extract an electron from $TiO_2$ at a terrace site to form an $(O_2^-)_{ads}$. Adsorbed $O_2$ species have a high binding energy (*20*) and are immobile in STM.

While $(O_2)_O$ do form on as-dosed surfaces with subsurface $V_O$'s, their concentration is low. The spontaneous healing process depicted in Fig. 1 happens rarely. $V_O$'s are more stable in the subsurface region than on the surface by ~0.4 eV (*10*). The activation energy for a $V_O$ to hop from the surface to the first subsurface layer ranges from 0.6 to 1.2 eV and the barrier for the reverse process is higher by ~ 0.4 eV (*11*). In the bulk $V_O$'s diffuse with a much lower barrier of ~ 0.2 eV; they thus tend to avoid the surface. While an adsorbed, negatively-charged $O_2$ reverses the energy balance (Fig. 1), the bulk $V_O$'s seldom come close enough to the selvedge for the healing process to take place.

We pulled $V_O$'s to the adsorbate-free surface with a sufficiently negative STM tip. Fig. S10 shows that this process is clearly field-induced. Likely the field reaches into the semiconductor (tip-induced band-bending) and pushes away electrons that are more or less localized around the $V_O$. This ionizes and destabilizes the vacancy, similar to the effect triggered by draining of electrons via localization at the $O_2^{2-}$ in the FPMD simulations. The threshold bias voltage for this process is 4.5 eV; the process becomes efficient at 5.2 V. The field-induced migration of intrinsic defects within $TiO_2$ is already utilized in novel memory devices, but the nature of the mobile species is controversial (*25*). Our results show that $V_O$'s in $TiO_2$ can be manipulated by high electric fields.

The tip-induced transformations of the adsorbed $O_2$ species are summarized in Fig. 3F. We propose that, initially, superoxide $(O_2^-)_{ads}$ forms at regular $Ti_{5c}$ sites. These are transformed into peroxide ions; the increase in band-bending observed after the first tip-induced conversion step suggests that the $O_2$ becomes more negatively charged. A similar, thermally activated transformation occurs if the sample is heated to 200 – 300 K (Fig. S11). DFT predicts similar adsorption configurations for $(O_2^-)_{ads}$ and $(O_2^{2-})_{ads}$, except for a slightly longer bond length in the peroxide ion (1.33 vs. 1.48 Å) (*20*). By hybrid functional calculations (*26*) we found a barrier of ~ 0.3 eV to transform $(O_2^-)_{ads}$ into the more stable $(O_2^{2-})_{ads}$ species. Once the $(O_2^{2-})_{ads}$ is formed, somewhat higher bias voltages (and/or prolonged exposure to the field) help the adsorbed molecule to merge with a subsurface $V_O$, resulting in the bridging dimer at a lattice site, $(O_2)_O$. For $(O_2)_{extr}$ near the Nb impurity, the situation is different. $(O_2)_{extr}$ is dissociated by the tip with a threshold voltage of 1.6 to 2.5 V, smaller than the $3.3 \pm 0.1$ V needed for the $(O_2)_{ads} \rightarrow (O_2)_O$ conversion. Each of the resulting adatoms merges with an $O_{2c}$ atom to form an $(O_2)_O$ species, consistent with the DFT predictions (*23*). The presence of the positively charged Nb impurity changes the energetics or (field-induced) kinetics of $V_O$ bulk-to-surface migration.

Our results clearly show the importance of subsurface O vacancies in $TiO_2$ anatase. They also exemplify how the electric field, e.g. from a STM tip, can be used as an effective tool to control the charge state of photocatalytically active species. The field can induce transformations of adsorbed $O_2$ that, by merging with a subsurface $V_O$, ultimately lead to the formation of the $(O_2)_O$ interstitial. This is the by far most stable $O_2$ species on the anatase surface. It is also an important intermediate in the photo-oxidation of water (*23*), suggesting it could contribute to the higher photocatalytic activity of anatase relative to rutile. Manipulating and controlling the $(O_2)_O$ interstitial might be thus a key to further the development of more active O-rich $TiO_2$ photocatalysts for water oxidation (*27*).

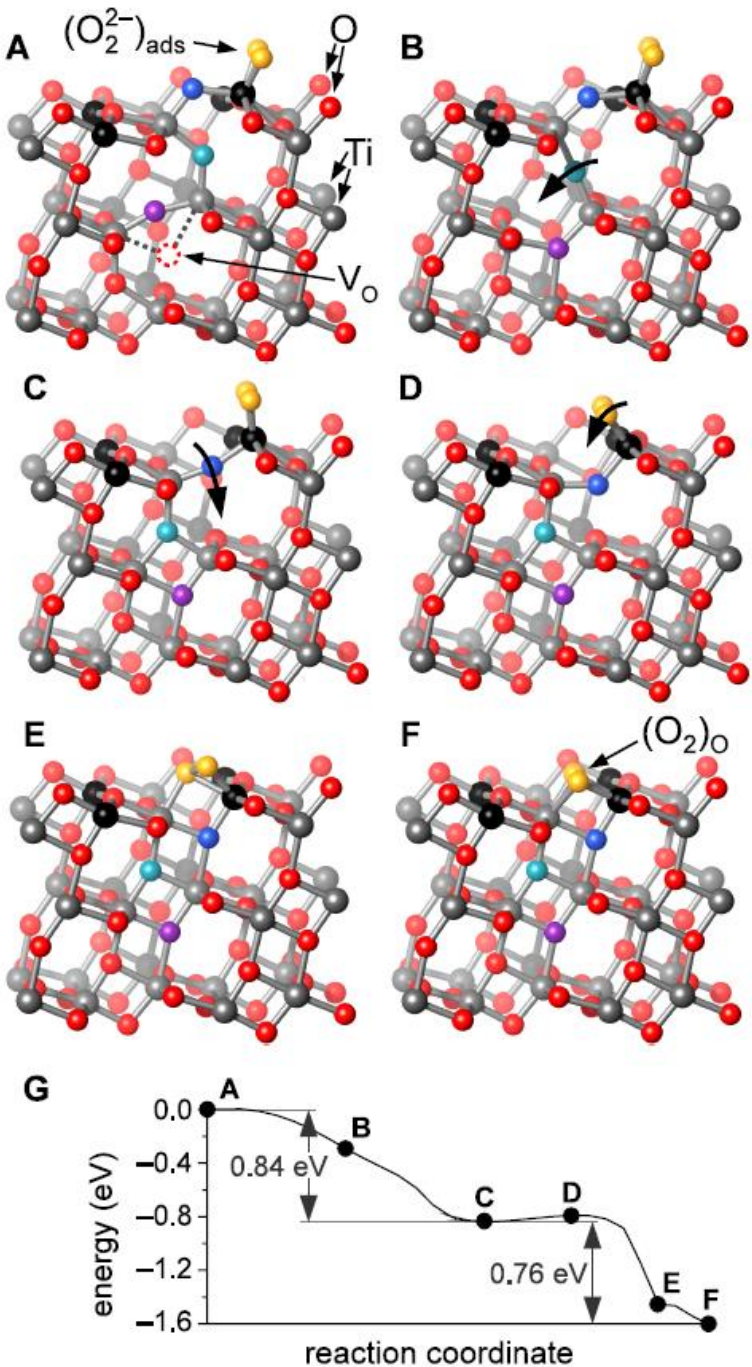

**Fig. 1 Reaction between an adsorbed peroxide, $O_2^{2-}$, and a subsurface oxygen vacancy at $TiO_2$ anatase (101).**
(A) Initially an $(O_2^{2-})_{ads}$ molecule (orange) is adsorbed at a 5-coordinated surface Ti atom (black). The O (violet) next to the subsurface $V_O$ is strongly relaxed toward the surface and returns to its lattice site in (B). (C) An O (cyan) from the bottom of the first $TiO_2$ layer fills the vacancy site, leading to a $V_O$ in an unstable position. In a final, combined event (D and E), the vacancy diffuses to the surface and is filled by the $O_2$ molecule. This chain of events heals the $V_O$ and the $O_2$ is incorporated into the surface as a bridging dimer, $(O_2)_O$ (F) The potential energy profile is shown in (G). See also Movie S1 (*12*).

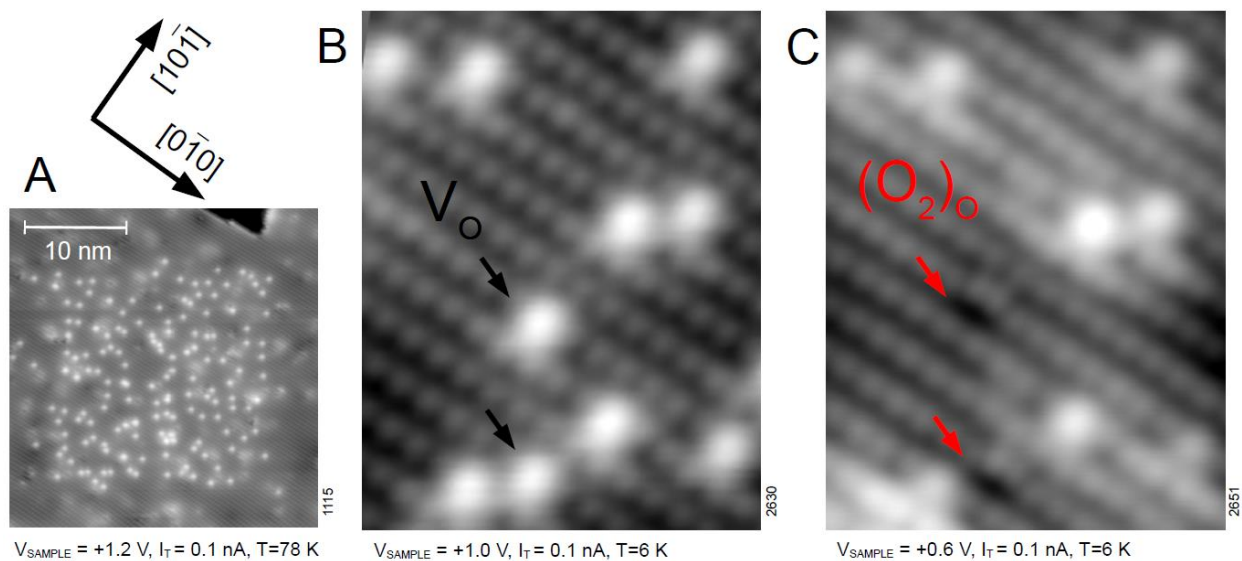

**Fig. 2 | Interaction of O₂ with surface oxygen vacancies (V$_O$) on TiO$_2$ anatase (101).** In reduced anatase, V$_O$'s are normally present within the bulk. In (A) these were pulled to the surface locally by scanning the center with a high sample bias of +5.2 V. In the high-resolution images of the same area before (B) and after (C) dosing 0.15 Langmuir O$_2$ at 45 K, the arrows point out V$_O$'s that reacted with O$_2$, forming the (O$_2$)$_O$ configuration shown in Fig. 1(F).

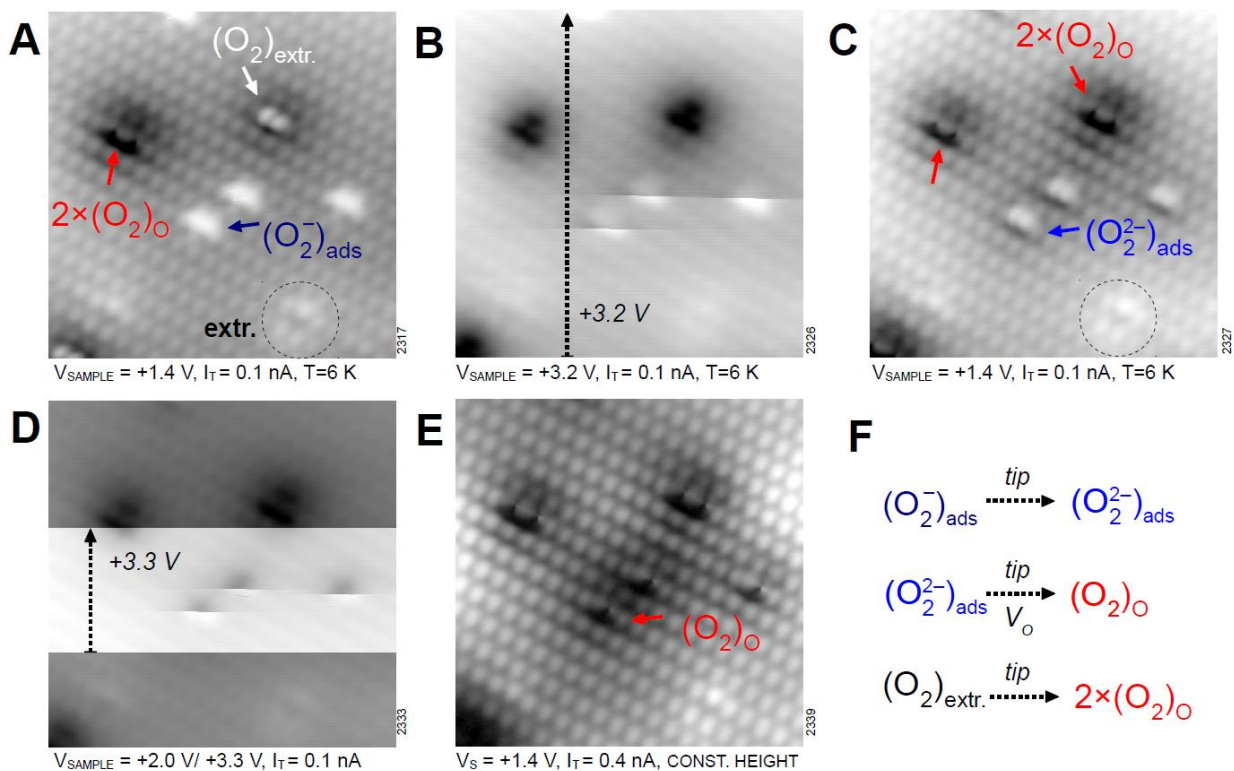

**Fig. 3 | STM tip-induced conversion of adsorbed O$_2$.** Sequence of STM images of the same area of an anatase (101) surface after exposure to 20 Langmuir O$_2$ at 105 K. (A) Various configurations of O$_2$: at regular sites, (O$_2^-$)$_{ads}$, at the position of lattice oxygen (O$_2$)$_O$, and at the dopants (O$_2$)$_{extr}$. The dopants are likely Nb, one of them is marked

with a dashed circle. After increasing the sample bias voltage to +3.2 V (B), the $(O_2)_{extr}$ is converted to two $(O_2)_O$ species, and the $(O_2^-)_{ads}$ to an intermediate species, likely a peroxo ion $(O_2^{2-})_{ads}$. (C) An additional scan with $V_{sample}$ = +3.3 V (D), converts each $(O_2^{2-})_{ads}$ into one $(O_2)_O$ (E). (F) Overview of the tip-induced processes.

**References and Notes:**

1. A. L. Linsebigler *et al. Chem. Rev.* **95**, 735 (1995).
2. M. A. Henderson. *Surf. Sci. Rep.*, **66**,185 (2011).
3. Z. Dohnálek *et al. Progr. Surf. Sci.* **85**, 161 (, 2010).
4. U. Diebold. *Surf. Sci. Rep.* **48**, 53 (2003).
5. S. Wendt *et al. Science* **320**, 1755 (2008).
6. G. Kimmel *et al. Phys. Rev. Lett.* **100**, 196102 (2008).
7. P. Scheiber *et al. Phys. Rev. Lett.* **105**, 216101 (2010).
8. H. Onishi *et al. Phys. Rev. Lett.* **76**, 791 (1996).
9. N. G. Petrik *et al. J. Phys. Chem. C* **115**, 152 (2010).
10. Y. He *et al. Phys. Rev. Lett.*, **102**, 106105 (2009).
11. P. Scheiber *et al. Phys. Rev. Lett.* **109**, 136103 (, 2012).
12. *See supplementary materials on Science Online*.
13. C. Pang *et al. Chem. Soc. Rev.* **37**, 2328 (2008).
14. X. Chen *et al. Chem. Rev.* **107**, 2891 (2007).
15. J. Augustynski. *Electrochim. Acta* **38**, 43 (1993).
16. A. Sclafani *et al. J. Phys. Chem. C* **100**, 13655 (1996).
17. Y. Nakaoka *et al. J. Photochem. Photobiology A* **110**, 299 (1997).
18. Y. Yamada *et al. Appl. Phys. Lett.* **101**, 133907 (2012).
19. M. Xu *et al. Phys. Rev. Lett.* **106**, 138302 (2011).
20. U. Aschauer *et al. Phys. Chem. Chem. Phys.* **12**, 12956 (2010).
21. U. Aschauer *et al. Proceedings SPIE*, 77580B (2010).
22. S. Na-Phattalung *et al. Phys. Rev. B* **73**, 125205 (, 2006).
23. Y.-F. Li *et al. J. Amer. Chem. Soc.* **132**, 13008 (2010).
24. O. Dulub *et al. Science* **317**, 1052 (2007).
25. K. Szot *et al. Nanotechnology* **22**, 254001 (2011).
26. Y.-F. Li *et al. J. Amer. Chem. Soc.* **135**, 9195 (2013).
27. V. Etacheri *et al. Adv. Functional Mater.* **21**, 3744 (2011).
28. P. Gianozzi *et al. J. Phys. Cond. Matt.* **21**, 395502 (2009)
29. J. Van de Vondele, *Comp. Phys. Comm.* **167**, 103 (2007)
30. Y. Furubayashi *et al. Appl. Phys. Lett.* **86**, 252101 (2005)

**Acknowledgments:** The experimental work was supported by the Austrian Science Fund (FWF; project F45) and the ERC Advanced Grant 'OxideSurfaces'. The theoretical work was supported by DoE-BES, Division of Chemical Sciences, Geosciences and Biosciences under Award DE-FG02-12ER16286. We used resources of the National Energy Research Scientific Computing Center (DoE Contract No. DE-AC02-05CH11231) and the TIGRESS high performance computer center at Princeton University.

**Supplementary Materials:**

Materials and Methods

Supplementary Text

Figures S1-S11

Tables S1-S2

Movies S1-S2

References (28-30)

**ERRATA:** "Reaction of $O_2$ with Subsurface Oxygen Vacancies on $TiO_2$ Anatase (101)" by M. Setvin (30. August 2013, p. 988). New results indicate that the species referred as $(O_2^-)_{ads}$ is an adsorbed water molecule, and the species named $(O_2^{2-})_{ads}$ is a terminal OH group. Revisited identification of the species is described in [Setvin *et al.*, Phys. Chem. Chem. Phys. **16**, 21524, 2014]. The field-induced oxygen vacancy migration remains valid as described in the Report; the process is discussed in detail in [Setvin *et al.*, Phys. Rev. B **91**, 195403, 2015]. All theoretical results, notably the pathway in Figure 1, remain valid as well.